\documentclass{article}

\usepackage{graphicx}
\usepackage{psfig}
\usepackage{epsfig}
\usepackage[round]{natbib}

\title{The problem with the Brier score}

\author{Stephen Jewson\footnote{\emph{Correspondence address}: RMS, 10 Eastcheap,
London, EC3M 1AJ, UK. Email: \texttt{x@stephenjewson.com}}\\
RMS, London, United Kingdom}
\begin{document}

\newcommand{\bx}[1]{\fbox{\begin{minipage}{15.8cm}#1\end{minipage}}}

\maketitle

\begin{abstract}
The Brier score is frequently used by meteorologists to measure the skill
of binary probabilistic forecasts. We show, however, that in simple 
idealised cases it gives counterintuitive results. We advocate the use of an alternative 
measure that has a more compelling intuitive justification.
\end{abstract}

\section{Introduction}

Users of meteorological forecasts need to be able to judge which forecasts are the best
in order to decide which to use. We distinguish two cases. The first case is one in which the user
plans to use the forecast for making a certain specific decision the details of which can be specified
entirely in advance. 
The second is one in which the user plans to use the forecast for making one or more decisions which
cannot be specified in detail in advance. 

In the first case it may be possible to decide which
forecast is the best by analysing the effect of using different forecasts on the quality of 
the final decisions made (for an example of this situation see~\citet{richardson00}).
In the second case, however, the user cannot convert forecasts into decisions ahead of time because they do not
know what decisions they are going to have to make. By the time they know what decision they are 
going to have to make, they do not have time to re-evaluate the available forecasts and potentially
switch to a different forecast provider. In this second case forecasts have to be analysed and compared
on their own merits, rather than on the merits of the decisions that can be based on them. 
In such a situation, the forecast user needs standard measures which can distinguish between forecasts at
a general level\footnote{A good example is the root mean square error, which is a general measure used
for comparing forecasts for the expectation}. It is this second case that we will consider. 

Forecasts can be divided into forecasts of the expectation of future outcomes
and probabilistic forecasts that give probabilities of different outcomes. 
Probabilistic forecasts can then be divided into continuous and discrete probabilistic forecasts.
A continuous probabilistic forecast gives a continuous density for the distribution
of possible outcomes. We have discussed how to 
measure the skill of such forecasts in~\citet{jewson03d} and have applied the measures we propose 
to the calibration and the comparison of forecasts in a
number of studies such as~\citet{jewsonbz03a} and~\citet{jewson03g}.

Discrete probabilistic forecasts give probabilities for a number of discrete events.
Any number of events can be considered, but in this article we will restrict ourselves 
to the case of only two events, which we call a binary probabilistic forecast.
We will address the question of how binary probabilistic forecasts can be compared.

One of the standard tools used by meteorologists to answer the question of which of two binary probabilistic
forecasts is the better is the Brier score, first used over 50 years ago~\citep{brier}, and still in
use today (see, for example, \citet{vitart03}, page 25). 
Nevertheless, we are going to argue that the Brier score is flawed. This is not something
that can be proven \emph{mathematically}, of course. Our arguments will be based on an appeal to 
intuition: we will present a simple case in which we believe it is intuitively clear which of two 
forecasts is the better, and we will show that the Brier score then comes to the opposite conclusion
to our intuition i{.}e{.} it gives the \emph{wrong answer}.
We will then present an alternative score that overcomes this problem, 
that has a definition that accords more clearly with intuition, 
and that is also more firmly grounded in standard statistical theory.

\section{The Brier Score}

The Brier score for a binary event is defined as:
\begin{equation}
  b=<(f-o)^2> 
\end{equation}
where $f$ is a forecast for the probability that an event $X$ will happen, and $o$ is an observation
which takes the value 1 if the event happens and 0 otherwise.
Lower values of the Brier score indicate better forecasts.
A detailed discussion is given in~\citet{tothet}.

We can expand the Brier score as:
\begin{equation}
  b=<f^2>-2<fo>+<o^2>
\end{equation}

When we are comparing two forecasts on the same observed data set the difference in the Brier score is given by:
\begin{equation}
  b_2-b_1=<f_2^2>-2<f_2o>-<f_1^2>+2<f_1o>
\end{equation}

where the $<o^2>$ term has cancelled because it is the same for both forecasts.
If this difference is positive ($b_2>b_1$) then we conclude that $b_1$ is the better forecast.

A particularly simple case is where the forecast probabilities have constant values, giving:
\begin{equation}
  b_2-b_1=f_2^2-2f_2<o>-f_1^2+2f_1<o>
\end{equation}

A further simplification is possible if the event occurs with a constant probability $p$, in which case $<o>=p$ and
\begin{equation}
  b_2-b_1=f_2^2-2f_2p-f_1^2+2f_1p
\end{equation}

\section{A simple example}

We now consider a very simple example, with constant event probability and constant forecast probabilities.
We set $p=\frac{1}{10}$, and consider the forecasts $f_1=0$ and $f_2=\frac{1}{4}$.

In this case the difference between the Brier scores is given by:
\begin{eqnarray}
  b_2-b_1&=&f_2^2-2f_2p-f_1^2+2f_1p\\\nonumber
         &=&\left(\frac{1}{4}\right)^2-2.\frac{1}{4}.\frac{1}{10}\\\nonumber
         &=&\frac{1}{16}-\frac{1}{20}\\\nonumber
         &=&\frac{1}{80}
\end{eqnarray}

The Brier score leads us to conclude that forecast $f_1$ is the better forecast.
However, this does not agree with our intuition. Forecast $f_1$ is a disaster: it predicts
a zero probability (a very strong statement!) for something that happens not infrequently.
Forecast $f_1$ is completely invalidated whenever event $X$ actually occurs (on average, 1
in every 10 trials).
Forecast $f_2$, on the other hand, is not so bad. It gives a lowish probability for something
that does indeed occur with a low probability. Its only fault is that the probability is
not exactly correct. 

The reason that the Brier score makes this mistake is that it does not penalise forecasts
that predict a zero probability strongly enough when they are wrong, even though our intuition
tells us that they should be heavily penalised. More generally, the Brier score does not
penalise forecasts that give very small probabilities when they should be giving larger probabilities
to the same extent that we penalise such forecasts with our intuition. This is because the Brier score is based on
a straight difference between $f$ and $o$.
Our intuition, on the other hand, considers the difference between probabilities of 0\% and
10\% to be very different from the difference between probabilities of 40\% and 50\%.
Intuition apparently uses fractional or logarithmic rather than absolute differences in probability.

One can easily construct other examples that illustrate this point. The more extreme the
events considered, the more striking is the problem with the Brier score. 
Consider, for example, $p=\frac{1}{1000}$, $f_1=0$ and $f_2=\frac{1}{400}$.
Again the Brier score prefers $f_1$, while our intuition considers $f_1$ to be a failure, and
$f_2$ to be a reasonably good attempt at estimating a very small probability.

We conclude that the Brier score cannot be trusted to make the right decision about which of
two forecasts is better. 
It should also not be used to calibrate forecasts or evaluate forecasting systems 
since it will over-encourage prediction of very small or zero probabilities.
We need a different measure.

\section{The likelihood score}

The standard measure used in classical statistics for testing which of two distributions gives the best
fit to data is the likelihood $L$ defined as the probability (or probability density) 
of the observations given the model and the parameters of the model~\citep{fisher1922}.
In our case this becomes the probability of the observations given the forecast.

We advocate the likelihood as the best metric for calibrating and comparing \emph{continuous} 
probabilistic forecasts (see the previous citations) mainly on the basis that it is very 
intuitively reasonable: the forecast that gives the highest probability for the observations
is the better forecast.
We also advocate the likelihood as the best metric for
calibrating and comparing binary forecasts. 
In this case the likelihood is given by:

\begin{equation}
    L=p(x|f)
\end{equation}

where $x$ is the full set of observations and $f$ is the full set
of forecasts. If we assume that the forecast errors are
independent in time then this becomes:
\begin{eqnarray}
L&=&\Pi_{i=1}^{i=n}p(x_i|f_i)\\\nonumber
 &=&\Pi_{i=1}^{i=n}o_if_i+(1-o_i)(1-f_i)
\end{eqnarray}
 We can also use the log-likelihood, which gives a more compressed
range of values, and is given by:
\begin{eqnarray}
    l&=&lnL\\\nonumber
     &=&ln [\Pi_{i=1}^{i=n}o_if_i+(1-o_i)(1-f_i)]\\\nonumber
     &=&\sum_{i=1}^{i=n}ln[o_if_i+(1-o_i)(1-f_i)]
\end{eqnarray}

If we put all cases of event $X$ occuring into set $A$, and all
cases of event $X$ not occuring into set $B$ then:
\begin{equation}
    L=\Pi_{A} f_i\Pi_{B} (1-f_i)
\end{equation}
and
\begin{equation}
    l=\sum_{A} f_i+\sum_{B} (1-f_i)
\end{equation}

If we now consider the special case in which $f$ is constant then:
\begin{equation}
    L= f_i^a (1-f_i)^b
\end{equation}
and
\begin{equation}
    l=a lnf_i+bln (1-f_i)
\end{equation}

where $a$ is the number of occurences of $X$, $b$ is the number of occurences of not $X$,
 and $b=n-a$.

If any of the predictions $f$ are 0 or 1 (i.e. are completely
certain) then $L=0$ and $l=-\infty$. If not, then $L>0$ and
$l>-\infty$. We see that use of the likelihood penalises the use
of probability forecasts with values of 0 or 1 very heavily.
Such forecasts get the worst possible score, as they should (since one can 
\emph{never} be completely certain).

In our simple example the difference in likelihoods for the two forecasts is:
\begin{equation}
  L_2-L_1=f_2^a (1-f_2)^b
\end{equation}

Since this is positive for all samples we see that the likelihood concludes that forecast 2 is better, in
line with our intuition.

\section{Summary}

Meteorologists have used the Brier score to compare binary probabilistic forecasts for over 50 years.
However, we find that in simple cases it makes the wrong decision as to which is the better
of two forecasts (where we define \emph{wrong} in terms of our intuition). We reach this conclusion
independently of any detailed analysis of the preferences of the user of the forecast. 

We advocate scores based on the likelihood as a replacement for the Brier score.
On the one hand the likelihood is conceptually simpler
than the Brier score: it decides which forecast is better simply according to which forecast gives
the higher probability for the observed data, which seems immediately reasonable. On the other
hand the likelihood accords with our intuition in the simple example that we present, and punishes
forecasts that give probabilities of 0 and 1 appropriately.

We conclude that use of the Brier score should be discontinued, and should be replaced by
a score based on the likelihood.

\section{Acknowledgements}

The author would like to thank Rie Kondo and Christine Ziehmann for useful discussions, 
and Christine for reading the manuscript and making some helpful comments.

\section{Legal statement}

The author was employed by RMS at the time that this article was written.

However, neither the research behind this article nor the writing of this
article were in the course of his employment,
(where 'in the course of his employment' is within the meaning of the Copyright, Designs and Patents Act 1988, Section 11),
nor were they in the course of his normal duties, or in the course of
duties falling outside his normal duties but specifically assigned to him
(where 'in the course of his normal duties' and 'in the course of duties
falling outside his normal duties' are within the meanings of the Patents Act 1977, Section 39).
Furthermore the article does not contain any proprietary information or
trade secrets of RMS.
As a result, the author is the owner of all the intellectual
property rights (including, but not limited to, copyright, moral rights,
design rights and rights to inventions) associated with and arising from
this article. The author reserves all these rights.
No-one may reproduce, store or transmit, in any form or by any
means, any part of this article without the author's prior written permission.
The moral rights of the author have been asserted.

\bibliography{jewson}

\end{document}